\documentclass[10pt,journal,cspaper,compsoc]{IEEEtran}

\usepackage[pdftex]{graphicx}
\usepackage{latexsym}
\usepackage{amsfonts}
\usepackage{amssymb}
\usepackage{amsmath}
\usepackage{epsfig}
\usepackage{epic}
\usepackage{eepic}
\usepackage{mathrsfs}
\usepackage{enumerate}
\usepackage{ifpdf}
\usepackage{multicol}
\usepackage{float}
\usepackage[T1]{fontenc}

\ifCLASSOPTIONcompsoc
\else
\fi

\ifCLASSINFOpdf
\else
\fi

\begin{document}
%
% paper title
% can use linebreaks \\ within to get better formatting as desired
\title{Bit Rate of Programs}

\author{Cewei~Cui,
        Zhe~Dang,
        and~Thomas~R.~Fischer% <-this % stops a space
\IEEEcompsocitemizethanks{\IEEEcompsocthanksitem C. Cui, Z. Dang, and T. R. Fischer are with School of Electrical Engineering and Computer Science, Washington State University, Pullman,
WA, 99164.\protect\\
% note need leading \protect in front of \\ to get a newline within \thanks as
% \\ is fragile and will error, could use \hfil\break instead.
E-mail: \{ccui,zdang,fischer\}@eecs.wsu.edu
%\IEEEcompsocthanksitem J. Doe and J. Doe are with Anonymous University.
}% <-this % stops a space
\thanks{}}

\IEEEcompsoctitleabstractindextext{%
\begin{abstract}
%\boldmath
A  program can be considered as a device that generates discrete time signals, where
a signal is an execution. Shannon information rate, or bit rate, of the signals may not be
uniformly distributed. When the program is  specified by a finite state transition system, algorithms are
provided in identifying information-rich components. For a black-box
program that has a partial
specification or does not even have a specification,
a bit rate signal and its spectrum are studied,
 which make use of data compression and the Fourier transform.
The signal provides a bit-rate coverage for testing the black-box
while its spectrum indicates a visual representation for execution's information characteristics.
\end{abstract}

\begin{keywords}
Shannon information, program, Lempel-Ziv compression, testing.
\end{keywords}}

% make the title area
\maketitle

\IEEEdisplaynotcompsoctitleabstractindextext
% \IEEEdisplaynotcompsoctitleabstractindextext has no effect when using
% compsoc under a non-conference mode.

% For peer review papers, you can put extra information on the cover
% page as needed:
% \ifCLASSOPTIONpeerreview
% \begin{center} \bfseries EDICS Category: 3-BBND \end{center}
% \fi
%
% For peerreview papers, this IEEEtran command inserts a page break and
% creates the second title. It will be ignored for other modes.
\IEEEpeerreviewmaketitle

\section{Bit rate of  programs}
A program consumes an input,  runs its instructions, and provides an output.
The input and
output can be encoded as strings. They are possibly
 interleaved (e.g., {\tt dollar?drink!dollar?drink!...} observed in a soft-drink vending machine).
An execution which is
a sequence of instructions, again, can be encoded as a string.
In this paper, we only consider programs that halt. That is, an execution is of finite but unbounded length.
When the program is deterministic,
there is only one execution on a given input. We consider an information theoretic
 \cite{shabook} model of a black-box  program as a channel:
\begin{itemize}
\item  On the sender side of the channel,  an input is fed into the channel;
\item  On the receiver side of the channel,  an execution along with the corresponding output is obtained.
\end{itemize}
A technically more convenient way is to treat the channel as a transducer $T$ that maps
a string $x$ (the input) to a string $y$ (an execution along with output on the execution. For now, we may ignore the output since it is
encoded in the output instructions in the execution. Hence, $y$ is simply an execution on input $x$.).
$T$, as a mapping, is many-to-many in general. When the program is deterministic,  $T$ must be many-to-one.

To avoid issues caused by granularity of executions, we understand an execution as a sequence of assembly instructions executed from
the compiled code of the program. In particular,  the input, as a string,
 is fed into the transducer/channel
 symbol by symbol (i.e., byte by byte), and therefore, the execution consumes the input also symbol by symbol (e.g., through {\tt
load-byte} instructions). We use $S_{\rm input}(n)$ to denote the number of
inputs with length $n$, and $S_{\rm exe}(n)$ to denote the number of executions with length $n$.
The {\em input bit-rate}  $\lambda_{\rm input}$ is defined as
\begin{equation}\label{inputrate}
\lambda_{\rm input}=\lim{{\log S_{\rm input}(n)}\over n}.
\end{equation}
When the limit does not exist, we take the upper limit, which always exists (for a
finite input alphabet). By convention, $\log 0=0$.  Throughout this paper, the  logarithm is base 2.
 Similarly, the {\em program (execution) bit-rate}  $\lambda_{\rm exe}$
 is defined as
\begin{equation}\label{programrate}
\lambda_{\rm exe}=\lim{{\log S_{\rm exe}(n)}\over n}.
\end{equation}
The notions used in (\ref{inputrate}) and (\ref{programrate}) come from
a well-known fundamental formula  proposed by Shannon \cite{shabook} in defining a
channel capacity which was  later
used by Chomsky and Miller \cite{chomsky58} for describing a complexity of regular languages.

\subsection{Intuitive explanations of the bit rates}
Intuitively,  the program bit rate, which is a nonnegative real number,  measures ``how many" executions are possible in the program.
Clearly, for nontrivial programs, the number of executions should be simply
 infinite ---- even when the program is deterministic, considering
the fact that the domain that the inputs are drawn from may be infinite.  To avoid this problem, the notion defined in
(\ref{programrate}) is used, which is always a finite number since the alphabet of instructions has
finite size. Theoretically,
the program bit rate $\lambda_{\rm exe}$ is exactly the number of bits per symbol needed when one losslessly compresses
an average execution. In other words,  a larger program bit rate makes the execution harder to compress, and therefore,
contains more information (hence, intuitively, the program's semantics is harder to comprehend).
In terms of (white-box) software testing,  its direct implication, according to the definition in
(\ref{programrate}),  is that the program is harder to test (i.e., ``more" execution paths to cover, even though the total number of paths
is infinite).

The input bit rate is similar. It characterizes ``how many" inputs could possibly be fed into the program.
Even though the number is usually infinite, the input bit rate, when calculated in (\ref{inputrate}) is always a finite number.
A larger input bit rate indicates that an average input carries more information per symbol. By looking at the
definition in (\ref{inputrate}),  it also says that,  in terms of (black-box)  software testing,
 the program is harder to test since the ``number" of inputs is also higher.

\subsection{Deterministic programs}
It can be shown that, for a deterministic program, the input bit rate is always greater or equal to
the program bit rate  (since an execution has to, as assumed, consume the entire input symbol by symbol, and there is only one
execution per input); i.e.,\footnote{
To show $\lambda_{\rm input} \ge \lambda_{\rm exe}$,
we notice that
$$|S_{\rm exe}(n)| \le \sum \limits_{i=0}^{n}|S_{\rm input}(i)|.$$
Hence,
$$\frac{\log{|S_{\rm exe}(n)|}}{n} \le \frac{\log{\sum \limits_{i=0}^{n} |S_{\rm input}(i)|}}{n}.$$
From $\lambda_{\rm input}=\lim {\frac{S_{\rm input}(i)}{i}}$,
for every $\epsilon>0$,
there is an $N_{\epsilon}$ such that, for each $i>N_{\epsilon}$, we have
$$\log{|S_{\rm input}(i)| \le (\epsilon+\lambda_{\rm input})\cdot i}.$$

\noindent Hence,
\begin{eqnarray*}
\frac{ \log{S_{\rm exe}(n)} }{ n } &\le& \frac{ \log{(\sum \limits_{i=0}^{N_{\epsilon}} |S_{\rm input}(i)| + \sum \limits_{i=N_{\epsilon}+1}^{n} |S_{\rm input}(i)|)} }{ n }\\
&\le& \frac{ \log{ (\sum \limits_{i=0}^{N_{\epsilon}} |S_{\rm input}(i)|  + n \cdot 2^{(\epsilon+\lambda_{\rm input})  \cdot n})}  } {n}.
\end{eqnarray*}

Taking $n \to \infty$, we have
$\lambda_{\rm exe} = \lim \frac{\log{|S_{\rm exe}(n)|}}{n} \le \epsilon + \lambda_{\rm input}$.

Sending $\epsilon \to 0$,
the result follows.
% we have
%$\lambda_{\rm exe} \le \lambda_{\rm input}$.
}
\begin{equation}\label{less}
\lambda_{\rm input}\ge \lambda_{\rm exe}.
\end{equation}
This observation is interesting in many ways.
Clearly, the inequality in (\ref{less}) says that an average input carries more information than an average execution.
In other words,  the program, understood as a channel,  adds redundancy to the information-rich
(i.e., harder to understand)  input so that the resulting execution is not so information-rich (i.e., easier to understand).
At an abstract level, a program is to solve a problem, where an input is simply an instance of the problem to solve;
this view can be found in any standard automata theory textbook. Therefore,  (\ref{less})  implies that, using
a deterministic program to solve the problem,  one {\em has to} ``stretch" an instance (i.e., the input)
 by inserting  redundancy and,
as a result, the bit rate of the solution (i.e., the execution)  is diluted and hence lower.  The essential reason of this comes from
the fact that a program, when compiled into assembly code can perform only extremely simple instructions per unit time.
This is evidenced by the simplicity of an instruction set of a modern processor that the program eventually runs on, where
its theoretical root comes from the equal (if not lower) simplicity of the instruction set in a Turing machine. Following this understanding,
it is clear that:
\begin{quote}
Deterministic programming adds redundancy to inputs. % which then  the processor can understand.
\end{quote}
Suppose that one tries to understand the semantics of the program (i.e., figure out that
the program indeed solves the problem that it intended to solve)  by tracing its executions.
Clearly, an execution represented at the assembly level has to be abstracted back to the source code
level  or even at the design level.  In terms of the channel we mentioned earlier,
it is a process of decoding an execution back to the input, which necessarily squeezes out all of the redundancy added by the program.
Hence,
\begin{quote}
Deterministic program understanding  removes redundancy in executions.
\end{quote}
What if (\ref{less}) does not hold? In this case, the program is simply not correct (does not solve the problem that it is intended to solve).
The conclusion is reached without any testing nor structural analysis on the code.

\subsection{Nondeterministic programs}
Nondeterminism is common in concurrent programming, which makes the transducer corresponding to a  program
be a many-to-many mapping. In other words, an input may result in multiple or even infinite number of executions.
Because of this,  the inequality  in (\ref{less}) does not always hold, and in some cases,
\begin{equation}\label{ge}
\lambda_{\rm input}\le \lambda_{\rm exe}.
\end{equation}
 In the light of the channel,  noise is added during transmission and thus,  on the receiver's side,
an execution is not only added with redundancy but also with noise. In the case when the information rate contained in the noise
exceeds the decrement of information rate caused by redundancy, (\ref{ge}) is therefore possible. A
nondeterministic program is known to be notoriously hard to design, develop, understand, and test.
One reason of this, in the view of information
theory, is that nondeterminism is interpreted as noise and hence makes the bit
rate of a program higher:
\begin{quote}
Nondeterministic programming adds redundancy {\em and} noise to inputs.
\end{quote}
A program with higher bit rate could be more information-efficient (per instruction, the execution carries more information) while,
on the other hand,
harder to understand (due to the same reason).\footnote{This is already known, for instance, in solving an NP-complete problem like
Boolean Satisfiability, where a nondeterministic algorithm only need ``guess" a Boolean assignment
 and check in linear time,  while a deterministic
algorithm is likely to enumerate all the possible assignments.}
\begin{quote}
Nondeterministic program understanding  removes redundancy {\em and noise} in executions.
\end{quote}
This likely implies, in practice,  a good program seeks a balance in the bit rate, which is not too high so that the code is
still understandable and not too low as well so that executions still carry a reasonable amount of information.

\subsection{Where are we heading?}
Given these information-theoretic understandings of programs,  the reader might bear the following doubt in mind:
would this help research in software engineering in certain ways?  The central  idea of the understandings defines
the notion of ``bit rate" (or a measure of information quantity)  of a program.  However, it is left unanswered where
the information is.  We think that an answer to this question is  a key to addressing the software
engineering aspects of the idea. We do not intend  to completely answer the question
in this paper. Instead, in the subsequent two sections, we probe the answers from the following two angles, respectively:
\begin{itemize}
\item  A program is a transition system,
 in particular, when one looks at a high-level design of the program.
Information may not be uniformly  distributed across the transition system.
So, when this transition  system
is treated as a white-box,  where is the information concentrated ?
\item  When the program is treated as a black-box,  its internal structure is not observable
(i.e., the transition system as well as its number of states is unknown).
However,
when it runs,  the execution (at the assembly level) can be observed.
On an execution, the bit rate may not be uniform: when the execution passes through an information-rich
component, the bit rate would be higher.
%Can we use the bit rate of the execution to infer some white-box properties of the black-box?
Recall that bit rate
of an execution is closely related to lossless compression  as we mentioned earlier. Therefore,
using an optimal universal compression algorithm (runs in linear time) like Lempel-Ziv in various ways,
we can obtain a bit rate signal of the execution. Using signal analysis techniques,
we can further use the spectrum of the signal to create  a  run-time ``coverage" of the blackbox.
\end{itemize}

\section{Bit rate of a program modeled as a finite state transition system}\label{section2}
%\subsection{An information rich component in a finite state transition system }\label{section3}
In this section, we consider a {\em finite state transition system}
$$M=\langle Q, R, q_{\rm enter}, q_{\rm exit}\rangle,$$
where $Q$ is a finite set of states, $R\subseteq Q\times Q$ is a set of transitions and with
$q_{\rm enter}\in Q$ being the entering state, and $q_{\rm exit}\in Q$ being the exit state.
We often write a transition as $q\to q'$ when $(q,q')\in R$.
For technical simplicity, herein, we assume that there is only one entering state and only one exit state;
when there are multiple entering states and exit states, the results in this section can be easily generalized.

For the given $M$, a {\em path} is a sequence of states $q_0\cdots q_n$, for some $n$, such that, for each $0\le i<n$,
$q_i\to q_{i+1}$.  In this case, the path is of length $n$, and
we say that $q_0$ {\em reaches} $q_n$, written $q_0\leadsto q_n$.
The information rate of $M$ is defined as
\begin{equation}\label{eqq0}
\lambda_M=\lim {{\log S_M(n)}\over n},
\end{equation}
where $S_M(n)$ is the number of paths, with length $n$,
 in $M$ from $q_{\rm enter}$ to $q_{\rm exit}$.
The rate can be efficiently and numerically
computed as the Perron number of
the adjacency matrix of the graph $M$ \cite{chomsky58}.
Indeed,
$M$ is essentially a directed graph.
Let $Q'\subseteq Q$.  We use $M'$ to denote the subgraph that only keeps nodes in $Q'$ and edges between
nodes in $Q'$ in $M$.
 As usual, $Q'$  is a strongly connected component
(SCC) if $Q'$ is maximal and satisfies,  for every $q,q'\in Q'$,  $q\leadsto q'$.

Notice that, in $M$,  a transition does not have a label. Doing this is purely due to technical convenience since
the information rates studied in a moment depend only on path counts. Of course, one may associate a label
with a transition $q\to q'$ in $M$; such a label, depending on the applications,  can be interpreted as, e.g.,
an I/O event.

Consider a path $\alpha$ from $q_{\rm enter}$ to $q_{\rm exit}$ in $M$.  As we mentioned earlier,  an information-theoretic
interpretation of $\lambda_M$ is the amount of information carried on $\alpha$ in terms of
average number of bits per symbol. However,  this amount of information is not necessarily uniformly distributed over $\alpha$;
some segment within $\alpha$ may carry more information; i.e., with higher information density. This is because
the path may pass through a subgraph that has a higher information rate. Identifying such a subgraph is practically meaningful,
since when $M$ is a design or even the code of a finite state program,  the subgraph (particularly when
containing most information)
 indicates a part of focus for testing. To do this, we need some more results.

Suppose that  the subgraph $M'$ is with its own
entering state $q'_{\rm enter}\in Q'$ and exit state $q'_{\rm exit}\in Q'$,
and  that the subgraph is reachable; i.e., $ q_{\rm enter}\leadsto  q'_{\rm enter}$ and
$q'_{\rm exit}\leadsto q_{\rm exit}$.  We first observe that
\begin{equation}\label{eqq1}
\lambda_{M'}\le \lambda_M;
\end{equation}
 i.e., the information rate of the subgraph is at most that of the entire graph $M$.
This can be shown as follows. Consider a path $\alpha$ that witnesses $ q_{\rm enter}\leadsto  q'_{\rm enter}$, and
a path $\beta$ that witnesses $q'_{\rm exit}\leadsto q_{\rm exit}$.
Clearly, for every path $\gamma$, from $q'_{\rm enter}$ to   $q'_{\rm exit}$,
with length $n$
 of the subgraph $M'$, the ``concatenated" path of $\alpha,\gamma,\beta$,  written $\alpha\gamma\beta$,  is also a path
of $M$ from $q_{\rm enter}$ to   $q_{\rm exit}$. This immediately gives
$$\lim {  {\log S_{M'}(n)}\over {|\alpha|+n+|\beta|}}\le \lim {
 {\log S_{M}(|\alpha|+n+|\beta|)}\over {|\alpha|+n+|\beta|}}\le\lambda_M.$$
 Noticing that the left hand side is
$$\lim {  {\log S_{M'}(n)}\over {|\alpha|+n+|\beta|}}=\lim {  {\log S_{M'}(n)}\over {n}} \cdot {n\over {|\alpha|+n+|\beta|}}=\lambda_{M'},$$
the result in (\ref{eqq1}) follows. (We should emphasize that the result is only asymptotic - generally doesn't
need to hold for finite $n$).

\subsection{Finding an information rich component}
Let $\theta$ be a number in  $[0,1]$.
From (\ref{eqq1}),  there exists a reachable subgraph $M'$, called a {\em $\theta$-information rich component ($\theta$-IRC)},
 which is minimal with
the rate $\lambda_{M'}\ge \theta\lambda_M$. Observe that a $\theta$-IRC is not necessarily unique.

Without loss of generality, let us assume that $M$ is cleaned; i.e.,
every node is on a path from the entering state to the exit state. Also we assume that $\lambda_{M}>0$
(otherwise, any subgraph is a $\theta$-IRC).
The following is a straightforward algorithm to find a $\theta$-IRC in $M$, which runs in worst-case
time $O(m\cdot {\rm Rate}(m))$,
where $m$ is the size (number of states + number of transitions) of $M$,   and
${\rm Rate}(m)$ is the time complexity, which is known efficient (polynomial time)
 in theory and in practice (as our experiments in \cite{rate10} implemented in MATLAB),
 of numerically computing the information rate in (\ref{eqq0})
 of a graph:

\vskip 0.25cm

{\bf Alg 2.1}

1.  Initially, every edge in $M$ is unmarked, and $M''=M$.

2.    If there is no unmarked edge, goto 7.

3. Delete an arbitrary unmarked edge  from $M''$  (but keep the nodes of the edge).

4.  Compute the information rate of the resulting  $M''$ ($M''$ is with the same entering and exist states as in $M$).

5.  If $\lambda_{M''}\ge \theta\lambda_M$,  then goto 2.

6.  Else (now the condition $\lambda_{M''}<\theta\lambda_M$ holds when $M''$ is without the deleted edge)
  add the deleted edge back to $M''$ and mark the edge, goto 2.

7.  (**) using Tarjan's algorithm finding a SCC in $M''$,  return the SCC (take  arbitrary
nodes in the SCC as the entering state and the exit state) as $M'$.

\vskip 0.25cm
We claim that the
SCC $M'$ returned from the algorithm is a $\theta$-IRC.  First observe that $\lambda_{M''}\ge\theta\lambda_M$
when in statement (**) a SCC is selected and marked as $M^\prime$, where $M'$ is a subgraph of $M''$. In particular, every edge is marked
in $M''$.  Suppose that $\lambda_{M'}<\lambda_{M''}$.
Let $\alpha$ be a simple path from the entering state to the exit state in $M'$.
The $\alpha$ can be obtained by deleting (at least one) edges $e$ from $M'$. After deleting these
edges,
$M''$ is a new graph, denoted by $\hat M''$.  Because $\lambda_{M'}<\lambda_{M''}$,
it can be shown that $\lambda_{\hat M''}=\lambda_{M''}\ge\theta\lambda_M$.
This violates the condition in line 5 of the algorithm when the edges $e$ were marked.
Due to the same reason,  $M'$ is minimal; i.e., dropping any edge from $M'$ will make it
with lower rate and hence be not a $\theta$-IRC anymore.

\begin{figure}[!t]
\centering
\includegraphics[width=7.5cm,height=5.5cm]{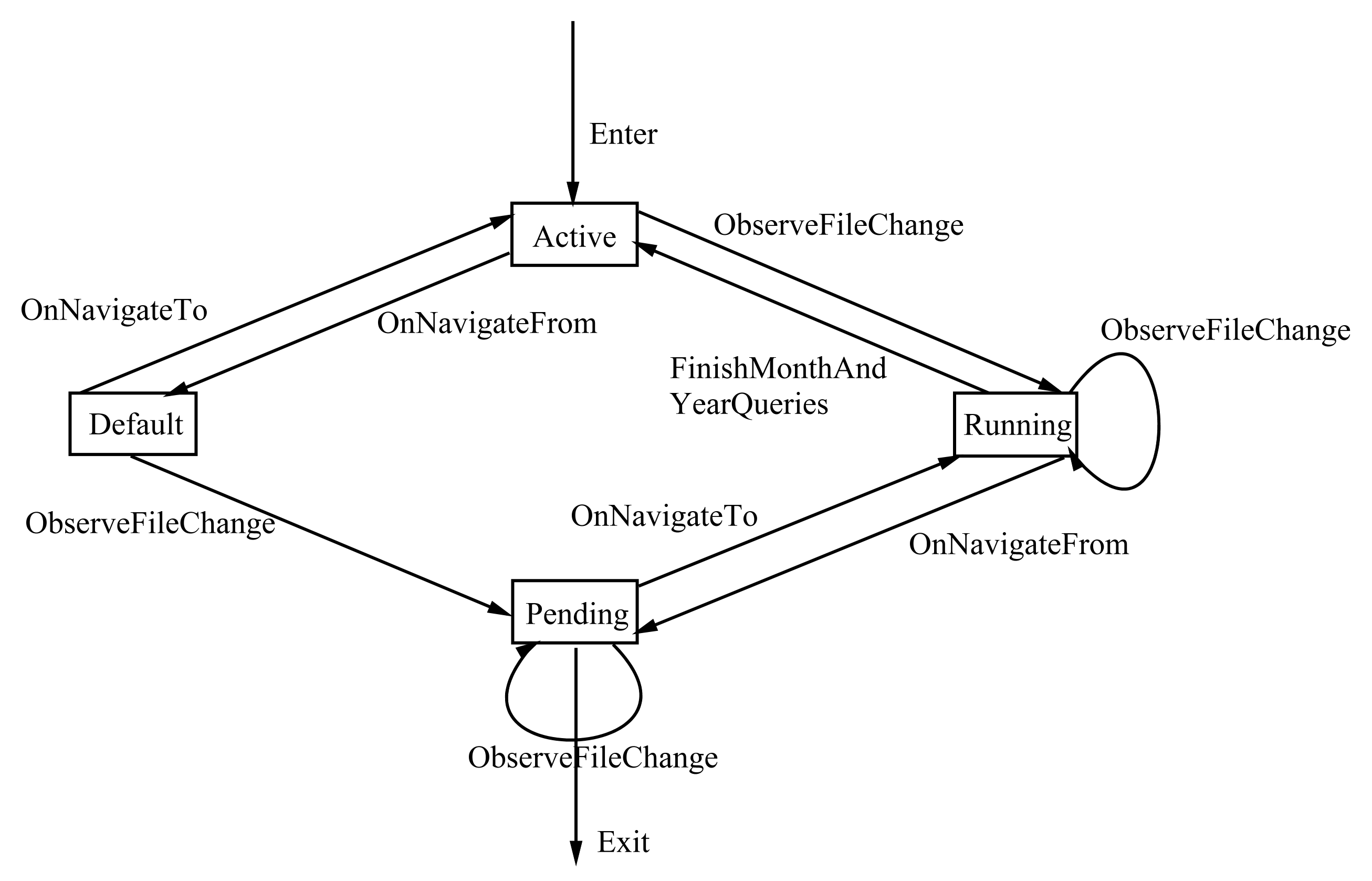}
\caption{A finite state transition system of Hilo's image browser view model \cite{Hilomodel}}
\label{thetaIRC}
\end{figure}
\begin{figure}[!t]
\centering
\includegraphics[width=7.5cm,height=5.5cm]{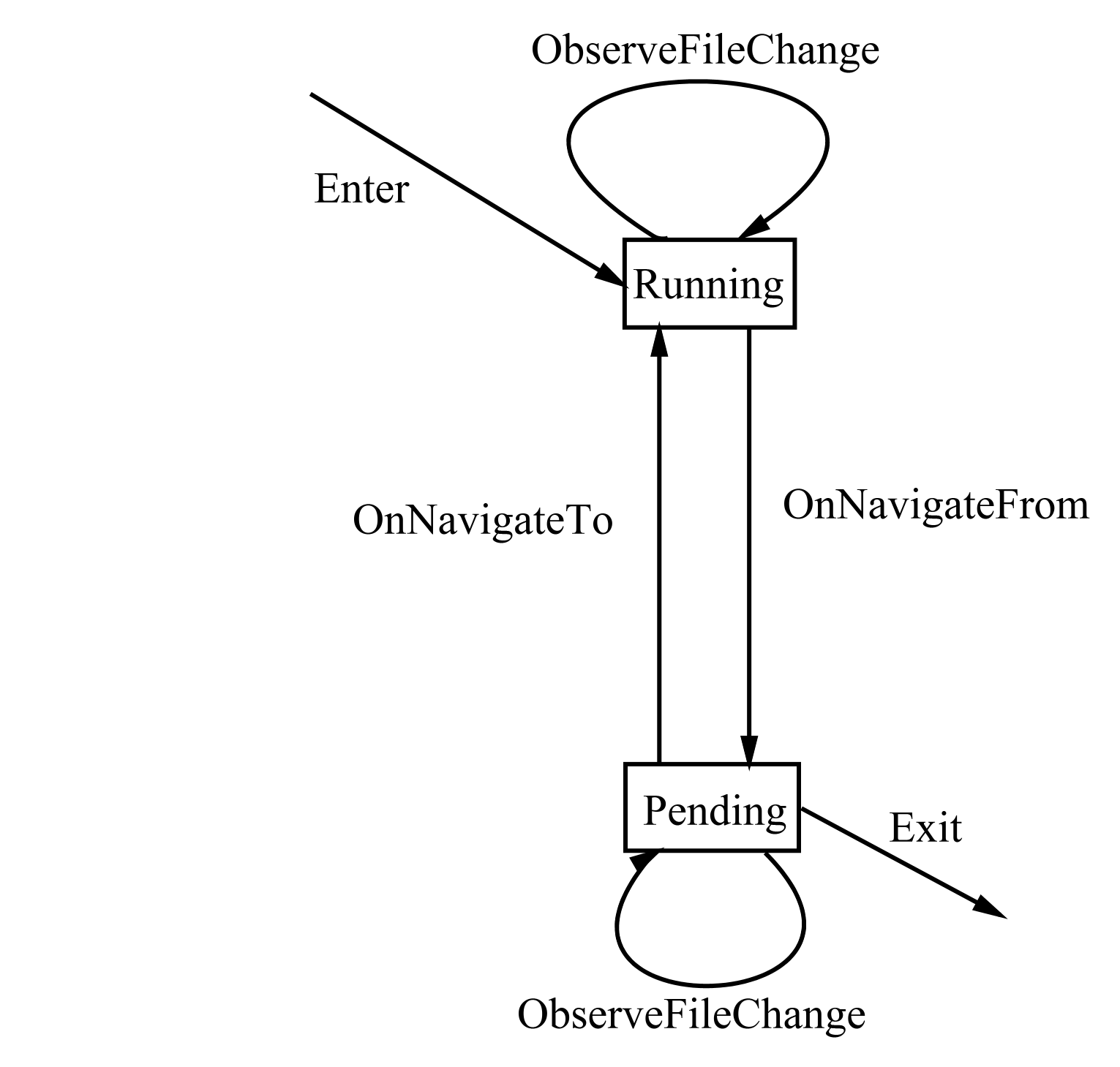}
\caption{A $\theta$-IRC of Figure \ref{thetaIRC} with $\theta$ being 0.79}
\label{delEdge}
\end{figure}

%An example is shown in Figure \ref{thetaIRC} and Figure \ref{delEdge}.
For example, Figure \ref{thetaIRC} shows a finite state transition system of Hilo's image browser view model \cite{Hilomodel}.
A $\theta$-IRC (with $\theta$ being 0.79) of the transition system
is shown in Figure \ref{delEdge}. That is, the IRC concentrates 79\% of the bit rate of the original
transition system.

%We can see that, compared to Figure \ref{thetaIRC}, Figure \ref{delEdge}
%only misses a self-loop for ``Pending" state such that its information-theoretic coverage is obviously reduced.

Now, consider two finite state transition systems $M_1$ and $M_2$ (with disjoint state sets).
As usual, we use $(M_1;M_2)$ to denote the sequential composition of the two (by connecting the exit state of
$M_1$ with the entering state of $M_2$). One can show that,  if $M_1'$  (respectively
$M_2'$) is a $\theta$-IRC of $M_1$ (respectively $M_2$), then $M_1'$ (respectively
$M_2'$)
  is also a $\theta$-IRC of the sequentially composed system
$(M_1;M_2)$.

The same result holds when $M_1$ and $M_2$ are composed nondeterministically as $(M_1\square M_2)$
(by creating a new state with a transition to the entering state of $M_1$ and a transition to the
entering state of $M_2$).

When $M_1$ and $M_2$ are  synchronously composed as $(M_1\| M_2)$,
finding a $\theta$-IRC in the composed system is more difficult and deserves investigation.
The IRC informs a tester which parts of $M_1$ and $M_2$ shall be a focus for intensive testing.
We start with definitions.

We pair some states in $M_1$ and some states in $M_2$ together, and put the pairs in a set $\Pi$.
Each such pair $(s^1,s^2)\in\Pi$ is called a synchronization pair, whose intended meaning is that,
when $M_1$ is at state $s^1$, and $M_2$ is at state $s^2$,  in the synchronized system $(M_1\| M_2)$,
both shall move together by each firing a transition at the same time.
Define $\hat Q_1$ (respectively $\hat Q_2$)
 to be the states of $M_1$  (respectively $M_2$)
that do not appear in any pair of $\Pi$.
Using the standard interleaving semantics of concurrency,
mathematically,  a {\em path of $(M_1\| M_2)$}
 is a word $\alpha$ in alphabet $\hat Q_1\cup \Pi\cup \hat Q_2$ satisfying:
\begin{itemize}
\item when deleting symbols in $\hat Q_2$ from $\alpha$,
and deleting symbols in $Q_2$ from every synchronized pair appearing in $\alpha$,
we obtain a path in $M_1$ from its entering state to its exit state; and,
\item when deleting symbols in $\hat Q_1$ from $\alpha$,
and deleting symbols in $Q_1$ from every synchronized pair appearing in $\alpha$,
we obtain a path in $M_2$ from its entering state to its exit state.
\end{itemize}
We use ${\rm Paths}(M_1\| M_2)$ to denote the set of all paths of $(M_1\| M_2)$.
${\rm Paths}(M_1\| M_2)$ is a regular language, which can be accepted by a DFA (that is
a graph, with $O(n_1n_2)$ nodes, where $n_1$ and $n_2$ are the numbers of nodes in $M_1$ and in $M_2$, respectively),
still denoted by $(M_1\| M_2)$.
Therefore, the information rate $\lambda_{(M_1\| M_2)}$
of the DFA can be computed efficiently. Notice that the rate $\lambda_{(M_1\| M_2)}$ could be much higher than
the rates for $M_1$ and $M_2$, because of the nondeterministic interleaving between states in $\hat Q_1$ and
states in $\hat Q_2$ in $\alpha$.

In order to define an IRC of $(M_1\| M_2)$, we need more definitions.  As before, we assume that $\lambda_{(M_1\| M_2)}>0$.
Consider a subgraph $M_1'$ and a subgraph $M_2'$ of $M_1$ and $M_2$, respectively.
Different from the previous definitions of IRCs,  we require now that $M_1'$ (respectively $M_2'$) contains the
entering state and the exit state of $M_1$ (respectively $M_2$). The difference comes from the fact that
state pairs in $\Pi$ must be synchronized on a path. We call $(M_1', M_2')$ a $\theta$-IRC of $(M_1\| M_2)$
if $\lambda_{(M_1'\| M_2')}\ge\theta\lambda_{(M_1\| M_2)}$ and both $M_1'$ and $M_2'$ are minimal.
A similar algorithm to find a $\theta$-IRC for $(M_1\| M_2)$ is as follows.

\vskip 0.25cm
{\bf Alg 2.2}

1.  Initially, every edge in $M_1$ and $M_2$ is unmarked, and $(M_1', M_2')=(M_1,M_2)$.

2.  If there is no unmarked edge in both $M_1'$ and $M_2'$,   return $(M_1', M_2')$.

3. Delete an arbitrary unmarked edge  from either $M_1'$ or $M_2'$  (but keep the nodes of the edge).

4.  Compute the information rate of the resulting  $(M_1'\| M_2')$.

5.  If $\lambda_{(M_1'\| M_2')}\ge\theta\lambda_{(M_1\| M_2)}$,  then goto 2.

6.  Else  put the deleted edge back  and mark the edge, goto 2.

\vskip 0.25cm
It is clear that the $(M_1', M_2')$ returned from the algorithm satisfies $\lambda_{(M_1'\| M_2')}\ge\theta\lambda_{(M_1\| M_2)}$.
In fact, it is also minimal. Otherwise, if one deletes an edge $e$
from, say $M_1'$ and still $\lambda_{(M_1'\| M_2')}\ge\theta\lambda_{(M_1\| M_2)}$
holds,  then this $e$ (being a marked edge) cannot be marked in step 6. Hence, the algorithm returns a $\theta$-IRC $(M_1', M_2')$.
The algorithm has worst-case time complexity  $O(m\cdot {\rm Rate}(O(m^2)))$, where $m$ is the maximal size of
  $M_1$ and $M_2$.

\subsection{Information rich inputs to a finite state transition system}\label{section4}
We now consider inputs to a finite state transition system $M$ defined earlier.
Let $\Sigma$ be  nonempty and finite alphabet.
It is necessary now to associate a label $a\in\Sigma$ to a transition in $M$; as a result,
the $M$ is a {\em labeled} finite state transition system.  Notice that some transitions
are labeled by null symbol ($\epsilon$) instead of a symbol in $\Sigma$
 to indicate, e.g.,
 it is an unobservable transition performing some internal actions.
For technical simplicity, we deliberately ignore output symbols. In fact, one may also
associate an output symbol on some of the transitions labeled with $\epsilon$, to indicate,
on a path of $M$,  an output sequence of symbols can be observed in response to
the input sequence of symbols fed along the path.  Adding such output symbols does not change
any definition or algorithms in this section and hence, our simplification is without loss
of generality.\footnote{When composition, which is not studied in Section \ref{section4}, is concerned, it is not generally a good idea
to ignore output symbols. This is particularly true
when the transition system is defined with a powerful semantics like in I/O automata  \cite{Lynch89anintroduction}.}

So now,  a path $\alpha$ is a state-symbol sequence $q_0a_0q_1a_1\cdots a_{n-1}q_n$, for some $n$,
where
$q_i{\stackrel{a_{i}}{\to}}q_{i+1}$ is a transition in $M$ (where $a\in\Sigma\cup\{\epsilon\}$), for each $0\le i<n$.
The input word $a_0\cdots a_{n}$ on the path is denoted by $w_\alpha$.
In this section, we will identify a ``minimal" set of inputs
   that causes a highest information rate of execution paths in $M$.
The set is called a set of {\em information rich inputs (IRI)}.
This is practically meaningful for black-box testing, where an input in such a subset can be intuitively
considered as one carrying the most
information  with respect to the transition system.

However, defining (not to say finding) an IRI is difficult, due to the fact that an IRI could be
infinite and we need a feasible way
to make it "minimal".  One way to define an IRI is as follows.
One can run {\bf Alg 2.1} on the graph of $M$ to identify a $\theta$-IRC in $M$.
The IRC, by definition, has the  information rate  $\lambda_{M'}\ge \theta\lambda_M$, and it is minimal.
The rate is measured on the paths   walking inside the IRC. Hence,  the input on each such path forms an IRI (from now on we call it
$\theta$-IRI).
There is a small problem here since a path in the IRC is not necessarily a complete path
(from the entering state to the exit state of the original $M$).
To fix this, we choose a simple path $\alpha$ from the entering state to a state $q$ in the IRC, and
a simple path $\beta$ from  a state $p$ in the IRC to the exit state.
We use $\gamma$  to denote a path inside the IRC from state $q$ to state $p$.
Then,  we use $w_{\alpha\gamma\beta}$ to denote the input word on the ``concatenation" of
$\alpha,\gamma,\beta$. We now put all such $w_{\alpha\gamma\beta}$ in a set, for all the $\gamma$'s
 ($\alpha$ and $\beta$ are fixed). The set is a subset of input words, and is defined as the $\theta$-IRI.
Clearly, the $\theta$-IRI is regular, and it is accepted by the finite automaton specified by the $\alpha$ (treated as a single-path graph),
sequentially composed  with the IRC, and then the $\beta$ (treated again as a single-path graph).
The whole computation takes the same worst case time as {\bf Alg 2.1}. This $\theta$-IRI obtained from $M$ is denoted by
$\theta$-IRI($M$) and will be used below.

 Suppose  now that an input $w$ is drawn from a domain specified by
a regular language  $L$ on alphabet $\Sigma$.  Such a domain is
used to restrict ``valid" input to feed into the system.
For instance, in an ATM banking system, multiple withdrawals for more than 300 dollars
within a day are not allowed in many locations.  This requirement makes some arbitrary deposit-withdraw
sequences invalid.
When this $L$ is given but  the finite state transition system is not given, we may still
ask a similar question: what would be a ``minimal" subset of $L$ that contains the most information in $L$?
Let $M$ be a DFA to accept $L$. Then,  the minimal subset that we are looking for can be defined to be
the $\theta$-IRI($M$) obtained in the preceding paragraph.
This $\theta$-IRI,  which will be used below, is denoted by $\theta$-IRI($L$).

It becomes rather complicated when we are given a finite state transition system $M$ as well as input language $L$.
The difficulty now is that an IRI in $L$ may not be the inputs carried on an IRC of $M$.
We first define an automata-theoretic construction
of an NFA $M'$ as follows.   $M'$ works on  a path of $M$.  While reading the input, $M'$ simulates
$M''$ by feeding every input symbol on the path into $M''$. Meanwhile, $M'$, by memorizing the graph $M$ (which is finite),
checks that the path is indeed a path of $M$, from the entering state to the exit state.
At the end of the path,  $M$ accepts if $M''$ enters its own accepting state.
We use $P(M')$ to denote the set of paths accepted by $M'$.
Clearly, $M'$ accepts exactly the paths of $M$ (from the entering state to the exit state) that will ``consume" an input from $L$.
 The number of states in $M'$ is $O(nm)$ where $n$ is the number of states in $M$ and $m$ is the number of states in $M''$.
We shall notice that the rate $\lambda_{P(M')}$ is the maximal rate $M$ running on input words drawn from $L$.
So, the desired  $\theta$-IRI will be a minimal set of input words such that when $M$ runs on these input words,
its rate is at least $\theta\lambda_{P(M')}$.
Since now $M'$ only runs on inputs from $L$,
the desired $\theta$-IRI is simply the $\theta$-IRI($P(M')$) defined in the previous paragraph.
But this is not right:  the inputs to $M'$ are paths (instead of input words in $L$). We need project those paths into
input words as follows.
First,  using the procedure in the last
paragraph, we find $\theta$-IRI($P(M')$).
Second,  since $\theta$-IRI($P(M')$) is regular (see the previous paragraph),  an NFA $M'''$ can be constructed to
``project" every path in $\theta$-IRI($P(M')$) to the input on the path.  The resulting set of inputs,
denoted by $\theta$-IRI($P(M'))\downarrow_\Sigma$
is the desired $\theta$-IRI, and can be specified by the $M'''$ whose size is at most $O(nm)$ , where $n$ is the size of
$M$ and $m$ is the size of $M''$. The entire process of finding the
desired IRI takes worst-case time  $O(nm\cdot {\rm Rate}(O(nm))).$

\section{Bit rate and spectrum of a black-box}
  In the previous section, bit rate  is measured over
a specification as a finite state transition system, and therefore, one may
select an ``information rich"
 test case from the specification. Notice that by  information is meant the information on the specification
instead of on the black-box, due to the fact that, in practice,
there is a gap between a specification and an implementation.
A specification is not the code; it is a hint of the code at best.
However, it is the code,  instead of the specification,
 that runs on a CPU and may fail.
Using specification-based testing,
one can select a number of test cases from the specification and run them  on a black-box. However, after running the test cases,
 the natural conclusion reached is that
one really does not know how much  information of the black-box (instead of the specification) has been covered.
But is this conclusion necessarily true?

If the conclusion is not true, then  the picture of black-box testing could change fundamentally  in several ways. For instance, one could
 re-evaluate test cases generated in specification-based testing and see whether they have covered enough  information of the unknown
code. If not, some  heuristic  approaches could be used to re-generate test cases so that more information will be covered. Our belief, justified by the experimental results presented in this section, is that the conclusion above is not true in general.
The heuristic approaches will also be applicable to guide random testing over a
black-box \cite{duran84,TsoukalasDN93,Ntafos98} so that test cases
generated are not only based on the specification but also based on the
information of the black-box.   This even works when the specification is  not complete or even not available.

\begin{figure}[!t]
\centering
\includegraphics[width=7.5cm,height=4.5cm]{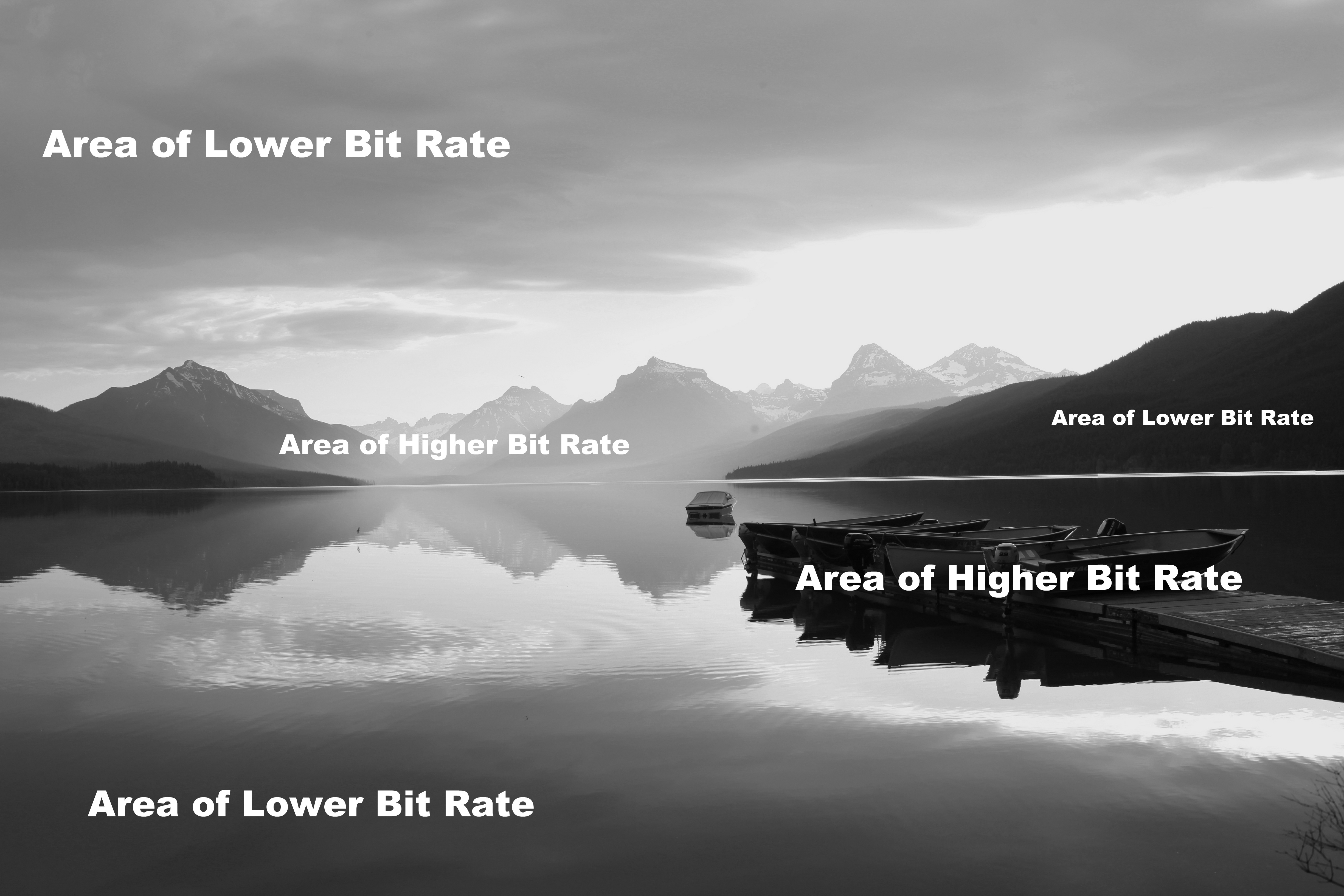}
\caption{A photo of Glacier National Park (U.S.), taken by one of the authors, showing that bit rate is not always
uniformly distributed and is often concentrated in several small areas.}\label{glaciers}
\end{figure}

As a motivating concept, consider the compression of a discrete-time signal. An effective compression algorithm, such as JPEG for images or Lempel-Ziv (LZ) for text, generates a bit-stream representative of the information rate in the signal. A signal with high (respectively, low) information rate is harder (easier) to compress. Common signals have non-uniform information rates as a function of time (or, for an image, spatial location). For example, Figure \ref{glaciers}
 shows a digital image where information (loosely, image activity) is concentrated in several relatively small areas. Hence, by monitoring the bit rate of the JPEG encoding bit-stream, one can roughly determine the information rate of different regions in the image. Furthermore, very efficient data compression algorithms, such as LZ or adaptive arithmetic codes \cite{RissanenL79}, include either implicit or explicit source modeling. That is, the encoding is performed conditioned on a context (of previously encountered and encoded source samples in the signal) and the required bit rate is reflective of the innovation in the signal, and hence, within the limits of the modeling, of the per-sample information rate of the signal.

Now, a black-box program can be viewed as a source device that emits an execution at run time, the execution corresponding to a sequence of assembly instructions. Analogous to the image source in Figure  \ref{glaciers}, the information rate of the execution carries valuable information about the source device (the program). An efficient data compression algorithm can be used to access this rate information, generating a signal to be analyzed and used to discover patterns in the software execution.

As before,  $S_{\rm exe}(n)$  denotes the number of executions (of the black-box under test)
 with length $n$.  The bit rate of the black-box is
defined, as before,
$$
\lambda_{\rm exe}=\lim {{\log S_{\rm exe}(n)}\over n}.
$$
However,  it cannot be computed since the code is not available. Despite this, $\lambda_{\rm exe}$ is
 exactly the average number of bits per symbol needed to losslessly encode
an average execution. That is, $\lambda_{\rm exe}$ is the average bit rate of an execution.
Every individual execution sequence has a bit rate, which indicates how much information is carried by the execution.
How can the bit rate of an execution be measured?
The Lempel-Ziv data compression algorithm \cite{ZivL77} (which is known to be universal)
 is one of the best and most commonly used compression algorithms, and
 is a good potential choice. Why?
Let's go back to the origin of this problem, and ask a broader question: what is the ``essential" information in an execution?
We may have different answers to this question from various perspectives.
One important perspective is provided by information theory. In information theory,
``non-essential" information is modeled as a form of statistical dependence, and is hence a form of redundancy.
The purpose of data compression (an application of information theory) is to reduce redundancy
for efficient representation. A sequence is more compressible
if it contains more redundancy.
Therefore, the rate  of a compressed sequence is an information-theoretic indicator that can measure information rate
in a sequence. %(A 99\% compression rate means almost not compressible.)

\subsection{Summary of our approach}
Our approach uses the following  two initial steps.
First,  in Lempel-Ziv encoding of an execution (for a given test case), a bit-stream is produced. By looking
up the dictionary formed in Lempel-Ziv encoding, the number of bits carried by each instruction, i.e.,
the {\em instantaneous bit rate} of the encoded execution sequence data, is obtained.
Through dividing the encoded execution sequence data into consecutive blocks,
we can calculate the average instantaneous bit rate for every consecutive block.
This yields a ``rate vs. time"
characteristic for the encoding.
Second, the rate vs. time characteristic is
some signal (here simply called the bit rate signal), and hence it can be analyzed using existing signal processing methods,
such as the Fourier transform, power spectrum estimation, smoothing filters, linear
prediction analysis, etc. Focusing on the Fourier transform, in this step a frequency domain spectrum of the signal is computed,  which is referred to as
the {\em (bit rate) spectrum} of the execution.

This approach is illustrated through an example.
 Consider a software system
{\tt bzip2} (a popular compression software package), which is
treated as a black-box. When the black-box
 runs under an input (test case, which is a file to be compressed),  we monitor its execution at the assembly level (this can be done by,
for instance, an instruction set monitor of the CPU or even in a debugger {\tt gdb}) and record a sequence of assembly
instructions, that is the execution.
Then, the sequence is compressed using Lempel-Ziv encoding
and the (instantaneous) bit rate of the sequence, (the compression length of each instruction in the sequence), is calculated.
%the sequence is parsed into consecutive segments of equal length and each segment is compressed using Lempel-Ziv encoding.
After that, the sequence is parsed into consecutive segments of equal length and
the average (instantaneous) bit rate of every segment is computed.
As a result, a time-domain
bit rate signal is generated, where the time index is the index of a block of  instructions.
Each segment serves, intuitively, as a ``rate region"  of the signal, analogous to the image
regions in Figure  \ref{glaciers}.

\begin{figure}[!t]
\centering
\includegraphics[width=6.5cm, height=5cm]{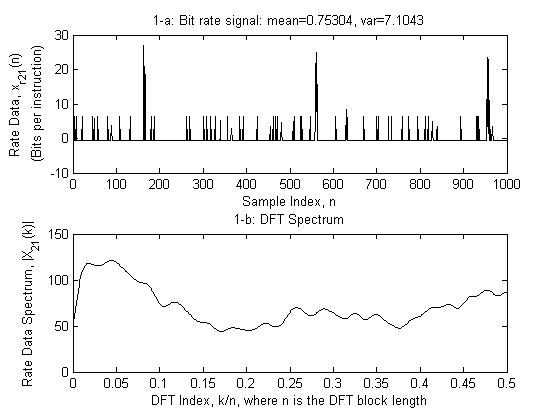}
\includegraphics[width=6.5cm, height=5cm]{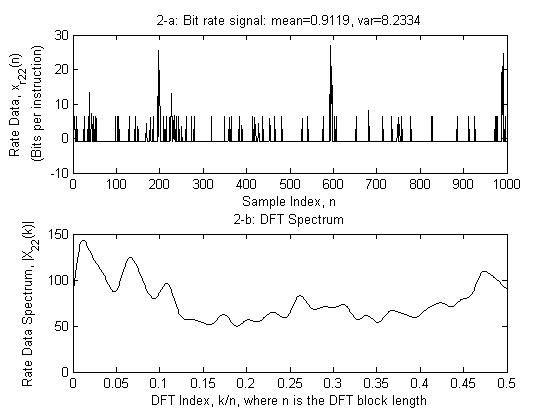}
\includegraphics[width=6.5cm, height=5cm]{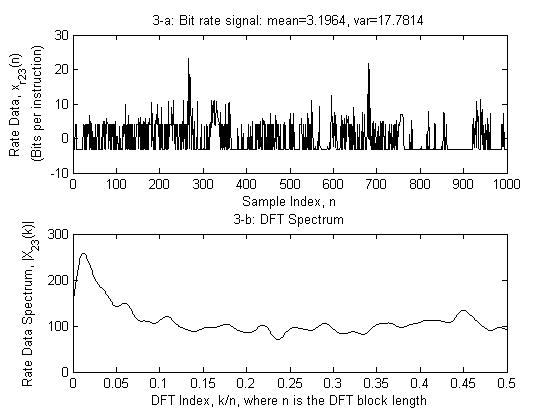}
\caption{
Three bit rate signals (1-a)(2-a)(3-a) for the same black-box under three test cases and
the three corresponding spectra (1-b)(2-b)(3-b) of the three executions, respectively.
 }\label{f12time}
\end{figure}

Figures \ref{f12time} (1-a)(2-a)(3-a) show three
bit rate signals corresponding to, respectively, three executions
 (for three different test cases, denoted as test-case-1, test-case-2, and test-case-3, which are
a PDF file, a WORD file, and a binary file) of the example black-box.
The length of each execution is 500,000 assembly instructions. For the figure, the execution instruction sequence is parsed into 1000 segments,
each  containing 500 assembly instructions.

The three bit rate signals look very irregular.
Signal processing techniques  can be used to apply a Fourier transform
so that the three signals
are, respectively,  transformed into frequency-domain bit rate spectra, as shown in Figure  \ref{f12time} (1-b)(2-b)(3-b).
(In order to smooth the spectra, a low-pass filter was also used.)
%since the curves generated by Fourier transform is quite noisy
%so that it is difficult to analyze a noisy curve.
%Although it is not easy not tell difference in appearance among \ref{f12time} (1-b)(2-b)(3-b),
%one may notice that amplitudes in these curves are quite different.
In the following, we define a distance between bit-rate signals and explain
the representation and meaning of the distance in time-domain and frequency-domain.

%INSERT
%----------------------------------------------------------------------
Let $x_r(n)$ be a bit-rate signal - that is, a sequence of non-negative bit-rates
for successively encoded blocks. The mean is $m_{x_r}=\frac{1}{N}\sum_{i=1}^{N} x_r(i)$,
and the mean-removed signal is $x(n)=x_r(n)-m_{x_r}$. (Note that
the bit-rate signals in (1-a),(2-a) and (3-a) are mean-removed signals.)
Let $X(k)$ denote the $N$-point discrete Fourier transform (DFT) of $x(n)$,
 where $x(n)$ is assumed to be of length $N$.
 Define the $\ell_2$ norm of a discrete signal, $x(n)$, as
 $$ ||x(n)||^2=\sum_{i=1}^{N}|x(i)|^2.$$
 Then the norms of $x(n)$ and its DFT, $X(k)$ are related by
 \begin{equation} \label{signal1}
 ||x(n)||^2= \frac{1}{N} ||X(k)||^2.
 \end{equation}
When comparing two signals $x(n)$ and $y(n)$,
the norm of the error signal, $x(n)-y(n)$ is given by $||x(n)-y(n)||$.
Let $x_r(n)$ and $y_r(n)$ be two bit-rate signals, with respective mean
$m_x$ and $m_y$, and define the mean-removed bit-rate signals as $x(n)=x_r(n)-m_x$
and $y(n)=y_r(n)-m_y$. Then the norms of the difference between original
bit-rate signals, and the mean-removed bit-rate signals, are related as
\begin{equation} \label{signal2}
|| x_r(n)-y_r(n)||^2 = || x(n) - y(n) ||^2 + N(m_x - m_y)^2,
\end{equation}
where $N$ is the length of the signals.

It is common to study the magnitude of a signal DFT, $|X(k)|$, and this
is often referred to as the (magnitude) spectrum. For power signals,
$|X(k)|^2$ is often referred to as the power spectrum. From (\ref{signal1})
the signal norm can be computed either from the time-domain signal,
or from its DFT spectrum.

Now, when comparing two DFT signals, say $X(k)$ and $Y(k)$, it is common
to plot their respective magnitude, $|X(k)|$ and $|Y(k)|$. However, the norm
of the difference of magnitude signals is not the same as the norm of the
difference of the two signals. That is, from (\ref{signal1}),
\begin{equation}
|| x(n) - y(n) ||^2 = \frac{1}{N}|| X(k)-Y(k) ||^2,
\end{equation}
and this is generally not $|| (|X(k)|-|Y(k)|)  ||^2/N $.
The operation of taking the magnitude of the DFT, namely,
$|X(k)|$, is generally information-lossy since
it eliminates all phase information.

As a result, we use (\ref{signal2}) to compute the distance.
 The corresponding
distances between (1-a) and (2-a), (1-a) and (3-a), (2-a) and (3-a)
are 15722.5, 28973.4 and 29331.2, respectively.
Clearly, the above results show that test-case-1 and test-case-2 are more similar
while test-case-1 (as well as test-case-2) is quite different from test-case-3.
(Notice that the norms of (1-a), (2-a) and (3-a) are only 7664.3, 9056.7 and 27980.8, respectively.
So the difference is significant.)
The distance, introduced in (\ref{signal2}) will yield a coverage indicator which will be described in detail in the following subsection.
In addition, we also show the mean and variance of bit rate signals in (1-a), (2-a) and (3-a), which can be
 a useful auxiliary indicator to differentiate distinct signals.
%-----------------------------------------------------------------------
%INSERT

We also need to clarify the significance of using the Fourier transform in our approach.
The distances between different signals
can be computed in either the time-domain or the frequency domain since the Fourier transform
is an invertible transform. It seems that everything can be done in the time-domain and the use of the
Fourier transform is redundant. Why do we choose to use the Fourier transform in this approach?
The main advantage of the Fourier transform is to provide an alternative, frequency domain, representation of a signal, revealing some signal characteristics not as readily apparent in the time domain.
%One classic example is human voice. In time domain, the voice of every person has similar wave forms.
%However, when Fourier transform being applied,
%the voices of males show significantly different frequency characteristics from the voices of females.
%In the spectra of males' voices, most peak values occur in low frequency part
%while peak values, in the spectra of females' voices, appear in high frequency part.
%This is consistent with the fact that voices of most adult male are with frequency 85$\sim$180 Hz \cite{voice1994}
%and voices of most adult female are with frequency 165$\sim$255 Hz.
One example is human speech. As a time signal, voiced speech exhibits quasi-periodic behavior,
while unvoiced speech appears noise-like.
In the Fourier domain, voiced speech exhibits clear spectral peaks,
with the pitch period of female speakers tending to be significantly
smaller (higher frequency) than for male speakers.
Additionally, the Fourier magnitude spectrum is time-shift invariant.
The Fourier transform is used in this paper to
provide an intuitive and visual representation of the frequency
domain information in an execution.

In addition, we shall point out that a bit rate spectrum can be obtained efficiently in
$O(n\log n)$ time from an execution (because the Lempel-Ziv algorithm is linear time and the (Fast) Fourier transform runs in $O(n\log n)$ time) and what we need is only an execution of the black-box instead of
its source code (which is assumed unavailable).
%The distance will yield a coverage indicator which will be described in detail in the following subsection.

\subsection{Experiments}
In this subsection, we will show a set of experiments to validate the usefulness of the bit rate spectrum  of a black-box program, and analyze its implications.

\subsubsection{Subjects}
In order to estimate and compare the bit rate of various programs, two classes of programs are chosen as subjects.
One class is  of large programs and the other class is of small programs.
The experiments consist of two groups.
Group 1 is designed to estimate the bit rate of a large program.
In the first group, bzip2 is chosen as the subject and 10 distinct inputs are fed to bzip2.
Group 2 is  designed to estimate the bit rate of a small program, selected from
a set of students' programming assignments, which implement PRIME (i.e., checking whether a  number is prime  or not).
Group 2 has   ten different inputs
 fed to the small program.
%It is known that the longer an execution sequence is, the more internal characteristics of a program can be shown.
%Therefore, for showing behavior characteristics of a program, sequence extracted from the execution of a program should be
%as long as possible.
In our tests, execution sequences for bzip2 were selected with length around 500,000;
 i.e. sequences of 500,000 assembly instructions are generated for bzip2.
However, considering the smaller size of PRIME programs, execution sequences of length around 100,000 were generated.

\subsubsection{Experimental Setting}
The procedure of our experiments is presented in the following.

First, using a {\tt gdb} script, we trace the execution of a program step by step (i.e. instruction by instruction) such that
a sequence of assembly instructions is generated.
In group 1, 10 different files, such as PDFs, pictures and binary executables, are fed to bzip2 as inputs so that
10 execution sequences are generated.
In group 2, one specific program implementation for PRIME  is chosen, and ten distinct inputs are fed to this implementation
 to generate 10 execution traces.

Second, we use the Lempel-Ziv algorithm to compress the execution trace.
Meanwhile, we look up the dictionary (at the instruction level) generated by Lempel-Ziv algorithm
to compute each instruction's compression length, i.e. instantaneous bit rate.

Third, each execution trace is equally cut into 1000 blocks, say $B_i, 1 \le i \le 1000$. Each block contains a sequence of consecutive instructions. The starting location of a block $B_{i+1}$ immediately follows the ending location of the previous block $B_i$. Using the instantaneous bit rate in previous step, the average
(instantaneous) bit rate of each block, (i.e., the summation of compression length of all instructions in the block divided by the number of instructions in the block),  is computed.
Then, using the Fourier transform and the low-pass filter, the bit rate  spectrum, which is a  frequency-domain signal,
is obtained.

\subsubsection{Results}
\begin{figure}[!t]
\centering
\includegraphics[width=6.5cm, height=4.5cm]{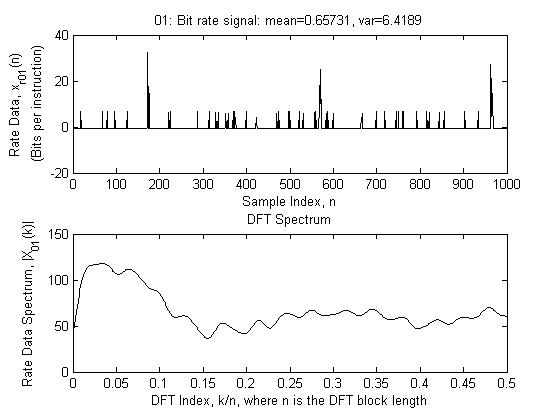}
\includegraphics[width=6.5cm, height=4.5cm]{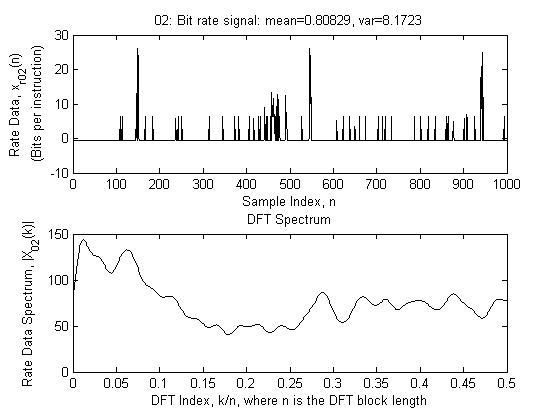}
\includegraphics[width=6.5cm, height=4.5cm]{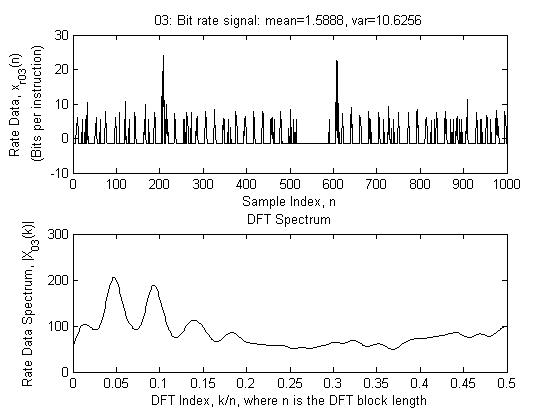}
\includegraphics[width=6.5cm, height=4.5cm]{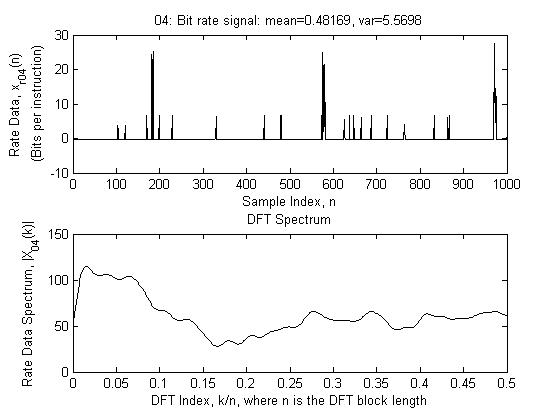}
\includegraphics[width=6.5cm, height=4.5cm]{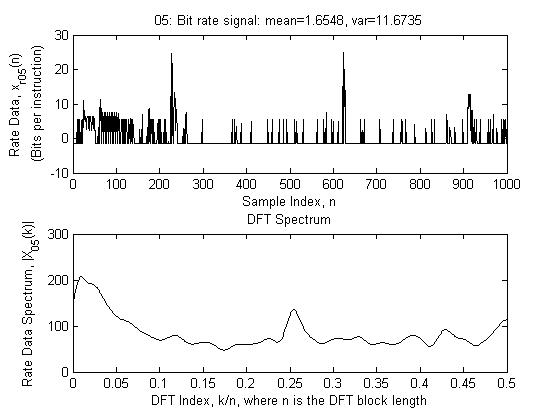}
\caption{
Cases 1-5 of group 1
 }\label{figure-g11}
\end{figure}

\begin{figure}[!t]
\centering
\includegraphics[width=6.5cm, height=4.5cm]{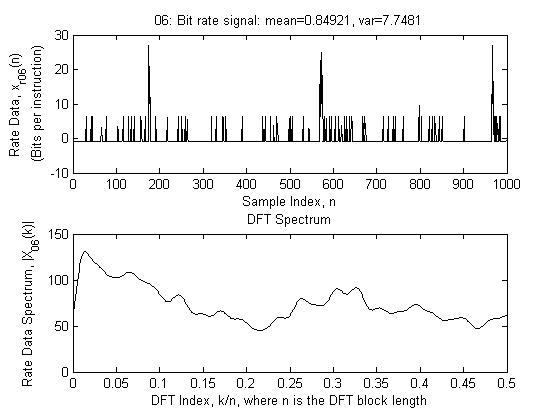}
\includegraphics[width=6.5cm, height=4.5cm]{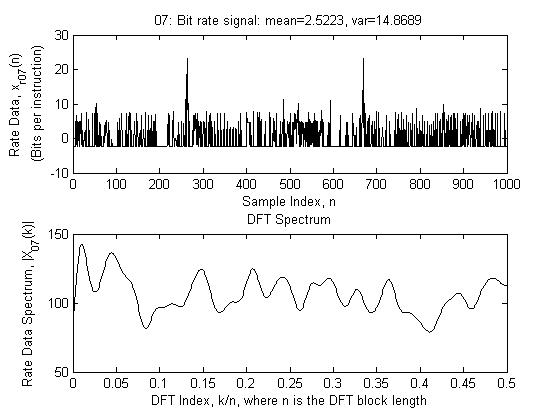}
\includegraphics[width=6.5cm, height=4.5cm]{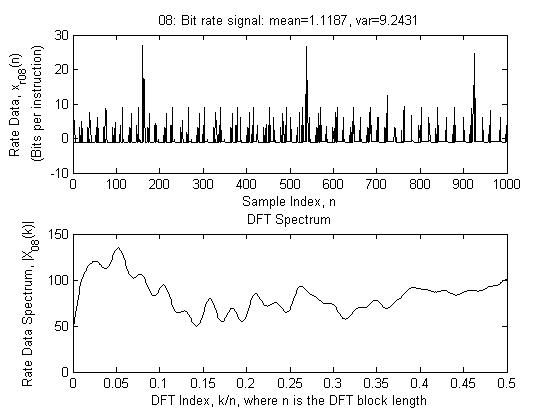}
\includegraphics[width=6.5cm, height=4.5cm]{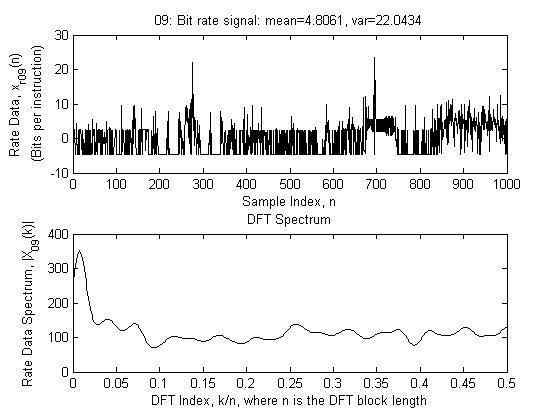}
\includegraphics[width=6.5cm, height=4.5cm]{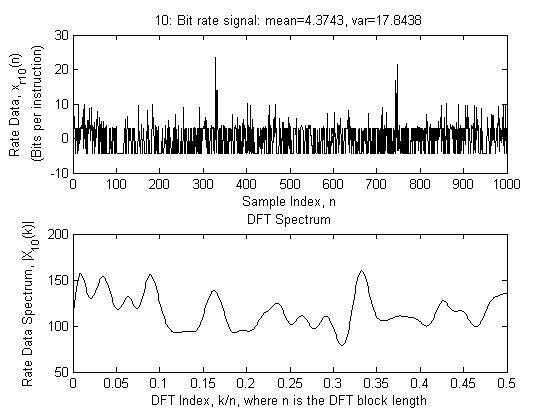}
\caption{
Cases 6-10 of group 1
 }\label{figure-g12}
\end{figure}

\begin{figure}[!t]
\centering
\includegraphics[width=6.5cm, height=4.5cm]{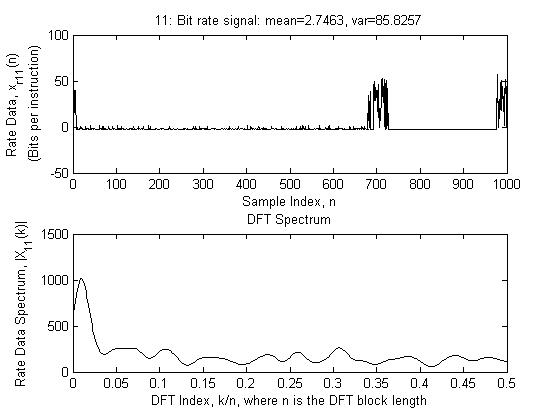}
\includegraphics[width=6.5cm, height=4.5cm]{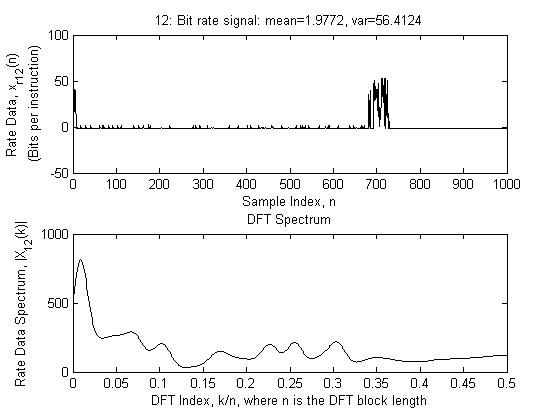}
\includegraphics[width=6.5cm, height=4.5cm]{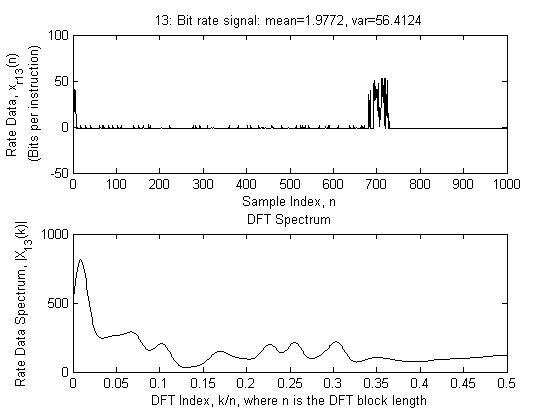}
\includegraphics[width=6.5cm, height=4.5cm]{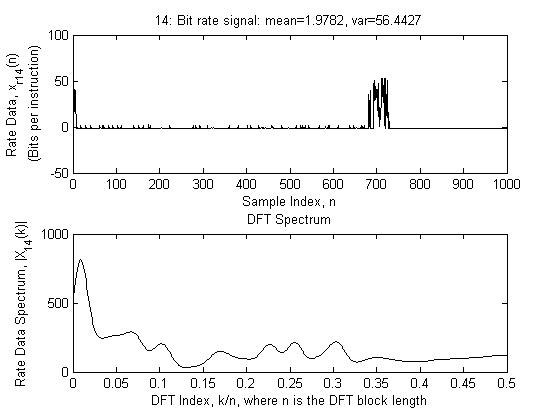}
\includegraphics[width=6.5cm, height=4.5cm]{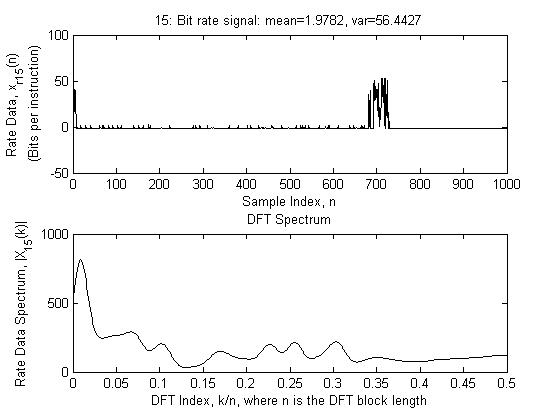}
\caption{
Cases 1-5 of group 2
 }\label{figure-g31}
\end{figure}

\begin{figure}[!t]
\centering
\includegraphics[width=6.5cm, height=4.5cm]{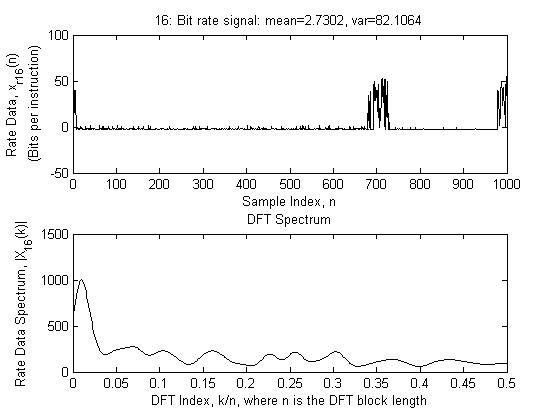}
\includegraphics[width=6.5cm, height=4.5cm]{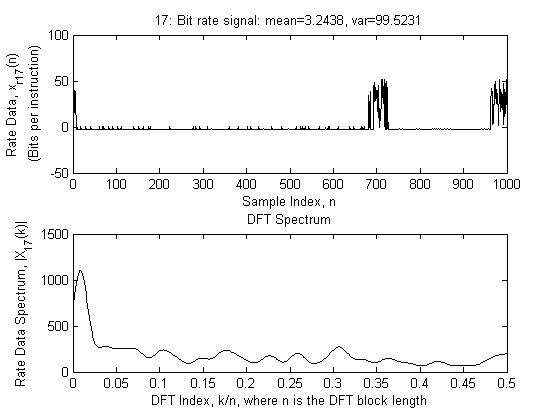}
\includegraphics[width=6.5cm, height=4.5cm]{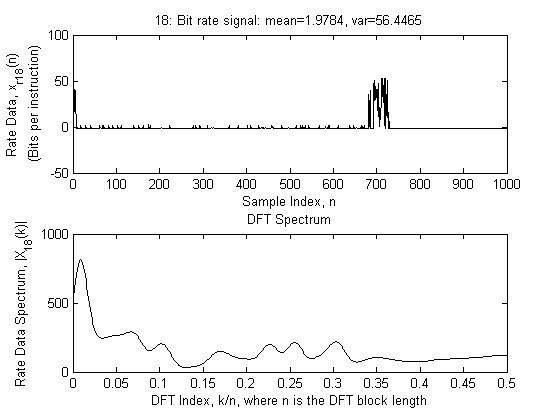}
\includegraphics[width=6.5cm, height=4.5cm]{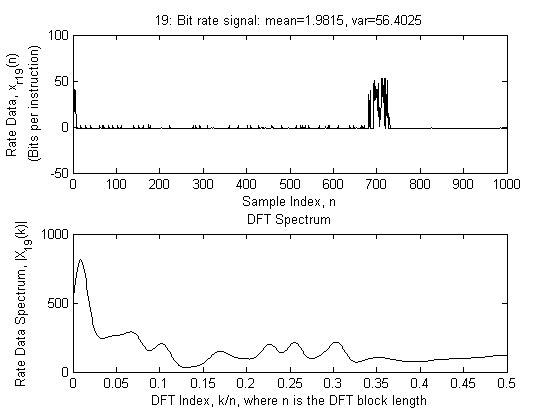}
\includegraphics[width=6.5cm, height=4.5cm]{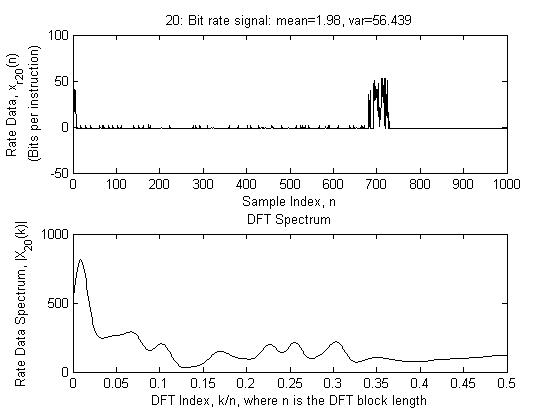}
\caption{
Cases 6-10 of group 2
 }\label{figure-g32}
\end{figure}

Figures \ref{figure-g11}-\ref{figure-g32} show the spectra for executions in groups 1 and 2.
We present our findings from the  results in the following.

1. Although it is difficult to compare bit rate signals in the time domain, it looks simpler to compare these signals in the frequency domain.
For instance, it is difficult to tell how different case 3 and case 5 in Figure \ref{figure-g11} are
 from their time domain bit-rate signals. However, it is easy to see the two signals show quite
different spectra in the frequency domain. Also, case 3 in Figure \ref{figure-g11} and case 6 in Figure \ref{figure-g12} have very similar looking time domain bit-rate signals, but
their spectra appear very different.

2. The spectrum reflects behavioral characteristics of a program. For example, case 3 in Figure \ref{figure-g11},
the magnitude spectrum has strong spectral peaks in the normalized
intervals [0.045,0.055] and [0.09,0.1]. What is the intuitive explanation of this observation?
Since the horizontal axis denotes frequency,
the large magnitudes at  a specific frequency implies some periodic tendencies in the bit-rate signal. We interpret this to mean that the information quantity of the execution has some roughly periodic aspects, and hence a concentration at certain
frequencies.  An almost flat spectrum indicates an almost uniform distribution of information along the execution, while
a spectrum with several peaks suggest an uneven distribution.
%It is known that, in physics, energy (or power) of a wave is proportional to the square of amplitude.
%If we say the process of a program's execution is the process of a program consuming its energy,
 %then the high magnitude frequency is also the part the program's energy concentrate.

3. Bit rate signals show a dynamic coverage of the black-box. Running a set of test cases on the same black-box,
we accordingly obtain a set of executions as well as their bit rate signals.
How much have the set of bit rate signals ``covered"?
 We mathematically define a
bit-rate coverage
as follows.  Intuitively, a set of two similar bit rate signals  should cover less than a set of two  bit rate signals that are not so similar.
In other words, one can treat a set of signals as a set of points in a metric space, and ask how much the points ``span"
in the space.  An ideal approach would be to use the Hausdorff content of the signal set. However,
since the dimension of the metric space is unknown in practice, this approach is not suitable.
Another approach is to borrow the idea of Borel cover by using  finitely many $\epsilon$-balls, in the metric space,
 to cover the points. Our preliminary studies show that it would end up with an exponential time
algorithm to find such a cover.
Herein, we propose a very intuitive and simple bit-rate coverage (indicator).
Let $T$ be a set of test cases and, for each $t\in T$, $x_{r\_t}(n)$ is the bit-rate signal of
the execution under test case $t$.
The bit-rate coverage
is defined as
$${\rm Cover}(T)={1\over 2}\sum_{t_1,t_2\in T} \| x_{r\_{t_1}}(n)-x_{r\_{t_2}}(n)\|^2.$$
For example, in group 2, 10 distinct inputs (test cases) are fed to the same program to generate
10 executions. The inputs of the first five cases are prime numbers while the inputs of
the  last five cases are non-prime numbers. The bit-rate coverage of the first five
cases is 136466.2 while
the bit-rate coverage of the last five cases is 274102.  The difference suggests that, obviously,
the first five test cases (all prime numbers) have covered less than the last five test cases (all non-prime numbers).
In other words, trying to extensively test the black-box, one should run more non-prime inputs than
prime inputs. Notice that, this conclusion is drawn without inspecting the code of the black-box.
This conclusion is consistent with our reading of the code:
the control flow of a prime input is much simpler than that of a non-prime input.

%Another example is from group 1. In group 1, 10 executions are generated by feeding 10 different
%inputs to bzip2. We divide all inputs into two subgroups: the first subgroup includes the first five cases
%in group1 and the second subgroup consists of the last five cases.
%The bit-rate coverage of the first subgroup is 40.1420 while the bit-rate coverage of the second subgroup is 44.5387. This fact indicates that the second subgroup covers more than the first subgroup. Inspecting the inputs. The first subgroup includes 4 text files and 1 picture file and the second subgroup consists 1 picture file and 4 binary files. In general, the process of compressing a text file is simpler than the process of compressing a binary file because text files, in general, contain less redundancy than binary files.

From the bit-rate coverage, we can also define a relative bit-rate coverage as follows:
$${\rm Cover}(t|T)=\sum_{t'\in T} \| x_{r\_{t'}}(n)-x_{r\_{t}}(n)\|^2,$$
where $t\not\in T\ne\emptyset$.
% Intuitively, for two test cases $t_1$ and $t_2$, when ${\rm Cov}(t_1|T)>{\rm Cov}(t_2|T)$,
%test case $t_1$ covers more information than test case $t_2$, for the given set $T$.
For instance, in  group 1,  we can calculate
$${\rm Cover}(\{{\rm case9}\}|\{{\rm case1,...,case8}\})=113564.3$$
and
$${\rm Cover}(\{{\rm case10}\}|\{{\rm case1,...,case8}\})=113938.2.$$
This intuitively says that, within the group,  case 9 covers almost the same additional amount of information as case 10.

\section{Discussions,  related work  and future work}
Why do we need to test software systems? Essentially, we test a
software system since there is uncertainty in its actual
behaviors (i.e., semantics). The uncertainty comes from the fact that
behaviors of
the software system are too hard to be analytically analyzed (e.g.,
the software system is Turing-complete), or even not available to
analyze (e.g., the software system under test is a black-box). In
other words, the actual behaviors  of the software
system are, at least partially,
 unknown. In our opinion, software
testing is an approach to resolve the uncertainty, and it gains
semantic knowledge of a software system by running it,
which resembles
opening the box to learn the situation of the famous
Schr\"{o}dinger's cat \cite{cat}, or more intuitively, the fact that
opening a box of chocolates resolves the uncertainty of what kinds of
chocolates are in the box.

How is ``uncertainty'' defined in mathematics?  Shannon entropy, or
simply entropy,
is specifically used to measure the amount of uncertainty in an
object in information theory \cite{sha48,cover06}, which is a
well-established mathematical theory underpinning all modern digital
communications. We have shown that, indeed, entropy in information theory can
be used to
characterize the uncertainty in a software system. More precisely,
black-box testing can be modeled as
a process of gaining information. We have developed  test case
selection algorithms that
obtain the most information (i.e., knowledge) and hence are
information optimal \cite{yang11}.
When a program is modeled as a finite state transition system
(represented as a labeled graph), its
information rate (or bit rate)
can be practically computed and, more importantly, the rate is
directed related to
the maximal entropy rate of a Markovian walk on the graph
\cite{rate10}. In the same paper, we also show,
through experiments, that the information rate of  C programs can be
estimated through its flow diagrams.
The information rate of a labeled graph can also be
used to select typical test cases \cite{cui}.
In our recent paper \cite{ewang}, we further  establish a relationship between
 the information rate of a program and the fault concentration when
the program is tested.
Automata are a universal model of all modern programs. Therefore, on
the fundamental side, we
have studied the information rate of automata and formal languages
\cite{rate_infinite,whitebox_rate,similarity}.
In this paper, we focus on information concentration within a program
(modeled as
a finite state system)  and on instantaneous information rate of an
execution, which,
as a result,  is a bit rate signal.

Our information-theoretic approaches
 to black-box testing
are  not intended to replace the existing specification
based testing techniques, such as FSM
\cite{Chow1978}, UIO \cite{Sabnani88},
testing hypothesis \cite{Gaudel95},
 Z-Specification \cite{Hierons97}, UML \cite{Briand01}, SCR
\cite{Heitmeyer96},
 and model checking based testing \cite{CGP99,HongCLSU03,GodefroidKS05},
 which is consistent with what the industry views a new testing approach.
Instead, our approaches will complement them.
Different testing approaches are just different angles from which
a tester looks at the system under test. We believe that our angle is
a new one.

In \cite{Chen2004},  Lempel-Ziv algorithms are used to detect plagiarism in programming assignments.
The use of Lempel-Ziv is to approximate Kolmogorov complexity, instead of Shannon information rate.
Additionally, source programs are compressed
in \cite{Chen2004} while, in our experiments,
execution sequences, i.e., run-time behaviors of source programs, are compressed.

What merits can our information-theoretic approach bring to
software testing? The most desirable property of Shannon entropy is that
the Shannon entropy
of a discrete random variable  remains unchanged after a one-to-one
function is applied \cite{cover06}. Such a characterization is of
great importance, since it suggests a way to test a software system based on
 its internal meanings (i.e., semantics), instead
of its appearance (i.e., syntax). For instance,
suppose that two distinct test sets are selected from a graph
modeling the system's control flow, both with 75\%, say, branch
coverage.  The adequacy degree does not differentiate the two sets.
Or, in other words, each branch is born equal. This is not
intuitively true: buttons on an LCD television do carry different
``amounts of information" (e.g., a television with a failed power-on
button is more ``useless'' than one with a bad volume button).

For the future work, many issues can be investigated.
For instance, blocks in this paper are of  the same
size. Averaging the bit rate over a block serves as a low pass
filter and hence a larger block size tends to
result in a smoother spectrum.
  In the  future, we need a study to determine advantages
and disadvantages of different block sizes.
Additionally, signal analysis techniques such as correlation and linear prediction
can also be used to analyze the internal structural information
of a black-box. The coverage developed in this paper
is based on the $\ell_2$ norm;  it is also worthwhile to develop other
distance measures; e.g., spectral magnitude based distances that are
invariant under signal time shift.

\begin{figure}
\centering
\includegraphics[width=7.5cm, height=5.0cm]{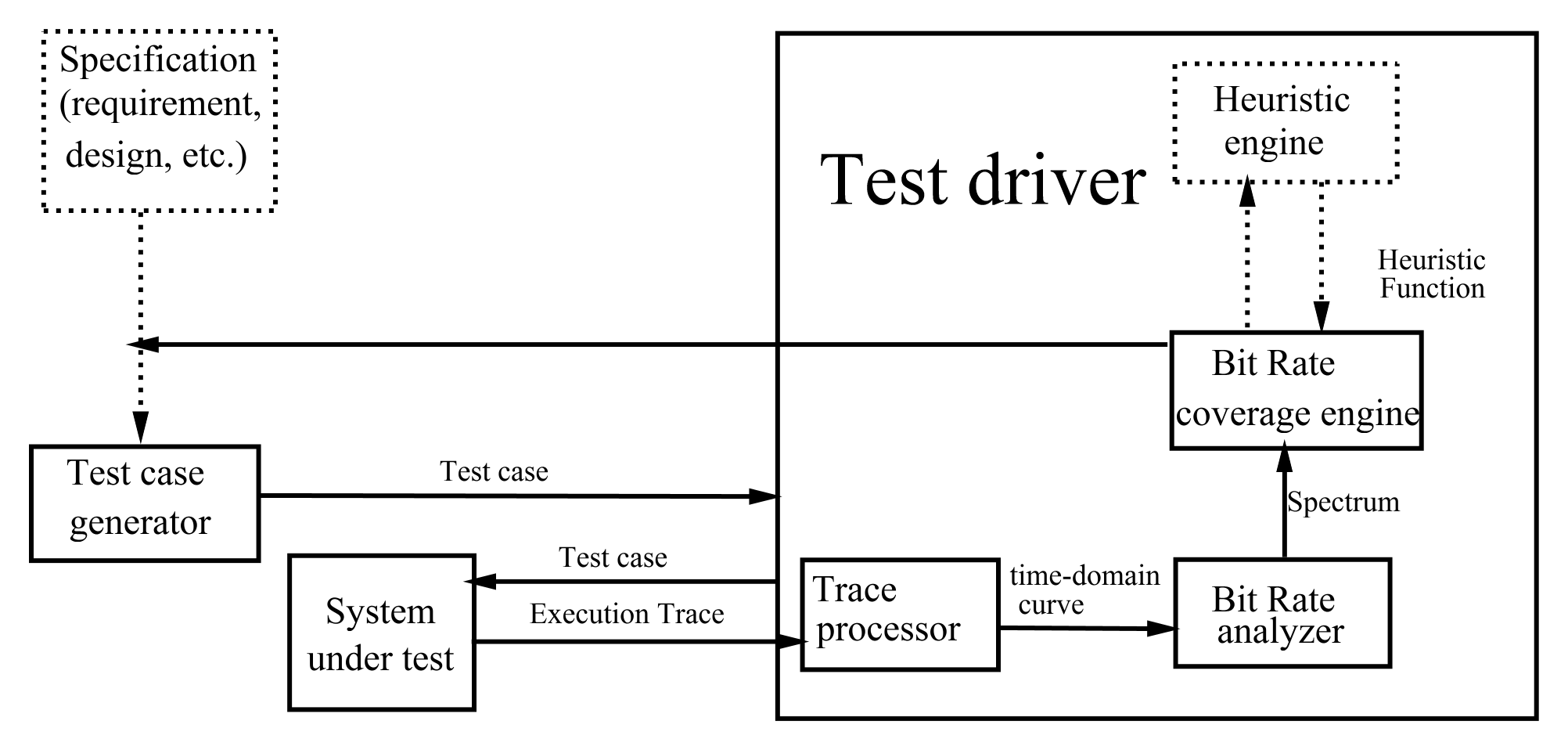}
\caption{
Framework of a spectrum black-box testing tool
 }\label{generator}
\end{figure}

Currently, the bit rate analysis approach can only work for evaluating
the bit-rate coverage of
a given set of test cases. It is not a test case generation approach.
To address the issue, we will
implement a testing tool whose framework is sketched in Figure \ref{generator}.
The testing framework mainly consists of three main components:
Specification (e.g., requirement, design, etc.; this component is
optional, i.e.,
the specification might be not available);
Test case generator (TCG) to generate  test cases
on the (black-box) system under test (SUT); and  Test driver that
analyzes and outputs results.

When the specification of the SUT is available,
TCG generates test cases according to the specification.
Once a test case is generated, the TCG feeds it to the test driver.
The test driver first runs the test case on the SUT while monitoring
run-time behaviors
(execution trace) of SUT.
If any fault occurs in the execution, the test driver will report and record it.
The trace processor parses the execution trace into a number of
segments such that
each segment contains an equal number of instructions.
Then, using the Lempel-Ziv algorithm,
the whole execution is compressed and the instantaneous bit rate of the execution
is obtained.
Hence, the average bit rate of each segment is calculated, yielding
an (abstract) time-domain bit rate signal.
The bit rate analyzer computes the discrete Fourier transform of the
bit rate signal,
producing  the bit rate spectrum (possibly smoothed using, for
instance, a low-pass filter).
The bit rate signal and spectrum are used by the coverage engine to
estimate how much bit-rate coverage has been achieved, for all the
test cases run.
The test driver can use the estimate  as a feedback to the test case
generator, and
 the feedback (combined with the  specification, if available) is used to
produce new test cases.
In the feedback step,  if the heuristic engine is available (and this
is to be developed in the project), some heuristic functions may
guide the process of producing feedback for the test case generator
such that the
test case generator selects a (possibly better) new test case.

Note that, in this framework, when the specification for SUT is not
available and
the heuristic engine is available,
it is an execution-based black-box random testing approach; when
neither specification nor
heuristic engine is available, it can be an execution based
evaluation framework on random testing. The
future research  includes study of how best to form the bit rate signal,
and which signal processing methods are most effective for extracting
meaningful patterns from the bit rate signal to estimate coverage.
In summary, our back-box testing approach, which is based on execution spectra,
aims to provide a practical testing approach to a significant problem
faced by the software industry:
 specifications are not always available,
and the system under test is complex enough that it needs to be
considered as a black-box.

% use section* for acknowledgement
\ifCLASSOPTIONcompsoc
  % The Computer Society usually uses the plural form

% Can use something like this to put references on a page
% by themselves when using endfloat and the captionsoff option.
\ifCLASSOPTIONcaptionsoff
  \newpage
\fi

\bibliographystyle{IEEEtran}
\bibliography{test,dang}

% that's all folks
\end{document}